\documentclass[review]{elsarticle}

\usepackage{float}
\usepackage{diagbox} 

\newsavebox\CBox
\def\textBF#1{\sbox\CBox{#1}\resizebox{\wd\CBox}{\ht\CBox}{\textbf{#1}}}

\usepackage{lineno,hyperref}
\usepackage{amsmath}
\usepackage{booktabs} 
\usepackage{multirow}
\usepackage{array}
\usepackage{amsmath}
\usepackage{subfig}
\usepackage{geometry}
\usepackage{booktabs} 
\usepackage{multirow}
\usepackage{threeparttable}
\usepackage{footnote}
\graphicspath{{figure/}}
\makeatletter  
\newif\if@restonecol  
\makeatother

\usepackage[linesnumbered,ruled,vlined]{algorithm2e}
\usepackage{algpseudocode}  
\usepackage{amsmath}  
\modulolinenumbers[5]  
\journal{**********·}
\begin{document}
\begin{frontmatter}
\title{Performance assessment and tuning of PID control using TLBO: the single-loop case and PI/P cascade case}

\author[mymainaddress]{Wei Zhang}
\author[mymainaddress]{He Dong\corref{mycorrespondingauthor}}
\cortext[mycorrespondingauthor]{Corresponding author}
\ead{donghe@hust.edu.cn}

\author[mymainaddress]{Yunlang Xu}
\author[mymainaddress]{Xiaoping Li}

\address[mymainaddress]{State Key Laboratory of Digital Manufacturing Equipment $\&$ Technology, Huazhong University of Science and Technology, Wuhan, 430074, China}

\begin{abstract}
Proportional-integral-derivative (PID) control, the most common control strategy in the industry, always suffers from health problems resulting from external disturbances, improper tuning, etc. Therefore, there have been many studies on control performance assessment (CPA) and optimal tuning. Minimum output variance (MOV) is used as a benchmark for CPA of PID, but it is difficult to be found due to the associated non-convex optimization problem. For the optimal tuning, many different objective functions have been proposed, but few consider the stochastic disturbance rejection. In this paper, a multi-objective function simultaneously considering integral of absolute error (IAE) and MOV is proposed to optimize PID for better disturbance rejection. The non-convex problem and multi-objective problem are solved by teaching-learning-based optimization (TLBO). This stochastic optimization algorithm can guarantee a tighter lower bound for MOV due to the excellent capability of local optima avoidance and needs less calculation time due to the low complexity. Furthermore, CPA and the tuning method are extended to the PI/P cascade case. The results of several numerical examples of CPA problems show that TLBO can generate better MOV than existing methods within one second on most examples. The simulation results of the tuning method applied to two temperature control systems reveal that the weight of the multi-objective function can compromise other performance criteria such as overshoot and settling time to improve the disturbance rejection. It also indicates that the tuning method can be utilized to multi-stage PID control strategy to resolve the contradiction between disturbance rejection and other performance criteria.
\end{abstract}

\begin{keyword}
 PID control\sep achievable performance\sep teaching-learning-based optimization\sep PID tuning\sep multi-objective optimization\sep temperature control.
\end{keyword}

\end{frontmatter} 

\section{Introduction}
In the last few decades, researchers have been working on the autonomous operation of industrial process control systems \cite{ozkan_advanced_2016,zhu_toward_2013}. PID control is the most widely used control method in the industry, and its performance maintenance has received much attention from both academia and industry \cite{gao_g081-novel_2017,gao_novel_2017}. 
Most of the process control systems suffer from performance deteriorating during long time operation under the influence of malfunction and environment such as equipment faults from sensor and actuator, controller tuning problem and changes of disturbance characteristic \cite{fang_g113-lqg_2017}. Therefore, there are many techniques to solve these problems, such as control performance assessment (CPA) \cite{yu_g045-performance_2016}, fault diagnosis \cite{yin_comparison_2012} and controller retuning \cite{veronesi_g112-performance_2009}. CPA aims to provide a benchmark for PID control systems to indicate the room for improvement, which plays a vital role in the autonomous operation of control systems \cite{byung-su_ko_g067-assessment_1998}. Once the model of the system and the characteristic of the environment are changed, a retuning process is needed to maintain the system performance.

Minimal output variance (MOV) is always used as a benchmark for CPA of PID control. However, MOV is not easy to be found due to the non-convexity of the relevant optimization problem. Therefore, many approaches have been proposed to solve this problem. Some approaches adopted local optimization methods \cite{byung-su_ko_g067-assessment_1998,agrawal_tuning_2003,ko_pid_2004} based on the gradient, but these methods can only provide an upper bound on MOV of PID since they do not ensure global optimality. Kariwala \cite{kariwala_fundamental_2007} reformulated the computation of MOV so as to ensure a lower bound. Sendjaja and Kariwala \cite{sendjaja_achievable_2009} represented the impulse response coefficients of the closed-loop transfer function as polynomials in unknown controller parameters and used sums of squares programming to solve the related optimization problem, which can guarantee a lower bound on the solution. Some researchers \cite{shahni_assessment_2011,veronesi_global_2011} employed global optimization methods to solve this non-convex problem to guarantee a lower bound. Nevertheless, since few of them analyze the calculation time, they are inappropriate to be applied in online CPA. Fu et al.\cite{fu_pid_2012} transformed the non-convex problem into a convex problem, which was solved by a low-complexity algorithm called iterative convex programming to promote the online application. To further reduce the calculation time, Shahni et al.\cite{shahni_rapid_2019} proposed a fast method by using a fixed length of impulse response coefficients to remove the iteration. This method with a weighting parameter can get a tighter lower bound, but the calculation time is longer. As stochastic optimization methods show great performance in solving global optimization problems, Pillay and Govender \cite{pillay_constrained_2020} proposed a hybrid algorithm combining Nelder-Mead simplex with Particle Swarm algorithm to solve this non-convex problem, but the results are not so competitive.

For the optimal tuning of PID controller, many objective functions have been proposed. Among them, integral of absolute error (IAE) and integral of the squared error (ISE) are the most commonly used criteria \cite{madhuranthakam_optimal_2008,tan_comparison_2006}, but they will lead to the contradiction between settling time and overshoot. Therefore, integral of time multiplied by absolute error (ITAE) and integral of squared time multiplied by squared error (ISTE) are proposed to overcome this problem \cite{krohling_2-design_2001}. However, these object functions can’t simultaneously optimize all criteria such as overshoot, rise time, settling time and steady error, and some multi-objective functions have been proposed recently to solve this problem \cite{sahib_1-new_2016}. Mouayad and Bestoun \cite{sahib_1-new_2016} proposed a new multi-objective function considering the four performance criteria simultaneously, and a decision making process is designed to select a best optimum from the Pareto optimal set. Zafer and Oguzhan \cite{bingul_4-novel_2018} proposed a novel multi-objective function taking into account mean of time weighted absolute error, settling time, overshoot and steady error. Zwe-Lee Gaing \cite{zwe-lee_gaing_3-particle_2004} proposed a new time domain performance criterion, the minimization of which corresponds to parameters with good step response. It also has literature that reports PID optimization with better disturbance rejection. Sigurd \cite{skogestad_tuning_2006} studied the tuning of smooth control systems for acceptable disturbance rejection, which is based on the simple Skogestad internal model control (SIMC) PID rule to provide the minimum limit of the gain. Renato and Joost \cite{krohling_2-design_2001} proposed to describe the disturbance rejection as ${H_\infty }$-norm, and then it is used as a constraint for the controller tuning with optimal disturbance rejection.

So far, few methods have the ability to solve the CPA problem of PID with accurate estimation and high efficiency simultaneously. To solve this problem, this paper proposes to use a stochastic optimization method named teaching-learning-based optimization (TLBO) that can balance these two aspects due to the capability of local optima avoidance and low complexity. Meanwhile, the algorithm is easy to implement and does not need any specific parameters \cite{rao_optimal_2015-1}. Although there are a lot of studies on the design of objective functions for PID tuning, few researchers consider the stochastic disturbance rejection. This paper proposes a new multi-objective function that simultaneously takes into account IAE and MOV. The weight of the function can compromise other performance criteria such as settling time and overshoot to improve the disturbance rejection of PID tuning. Furthermore, the optimal tuning and CPA are extended to the PI/P cascade control case, which is a practical control strategy in process control systems. Several simulation examples from the literature \cite{shahni_assessment_2011} are tested to demonstrate the superiority of TLBO in solving the CPA problem of PID. The results indicate that TLBO can obtain better MOV and runtime than existing methods on most problems. The proposed tuning method is applied to two temperature control systems, and the simulation results show that the method can improve the disturbance rejection to a large extent. Moreover, the method can combine with a multi-stage PID control strategy \cite{li_high_2017} to resolve the contradiction between disturbance rejection and other performance criteria such as overshoot and settling time.

This paper is organized as follows: Section 2 introduces the TLBO algorithm. Section 3 described the achievable performance and tuning method of PID control. In section 4, the results of the simulation examples are presented. Finally, section 5 shows the conclusion.

\section{Teaching-learning-based optimization}
The TLBO algorithm, proposed by Rao et al. \cite{rao_teachinglearning-based_2011} in 2011, has been a powerful meta-heuristic optimization algorithm in solving engineering problems due to the two-phase strategy, i.e., teacher phase and learner phase. This strategy imitates the process that learners improve their knowledge through teaching and learning behaviors. The algorithm involves two populations named learners and teachers, and the learner with the best fitness value at every iteration is chosen as the teacher. In the teacher phase, learners learn knowledge from the teacher to approach the global optimum, which guarantees the exploitation capability of the algorithm. In the learner phase, the learners learn knowledge from each other to get more chance to find the global optimum, which make the algorithm have excellent exploration capability.

\subsection{Teacher phase}
In this phase, the teacher tries to improve the mean of all learners at any iteration $ G $. Supposing that the number of learners is $ Np $ (for any learner $ i $, $ i=1, 2 , …, Np $) and the dimension of a learner is $ D $ (for any dimension $ j $, $ j=1, 2, …, D $). The learners can be updated by the following law
\begin{equation}
P_{i,j}^{(1)} = {P_{i,j}} + {r_i}({P_{teacher,j}} - {T_F}{M_j})
\label{eq1}
\end{equation}
where $ {P_{i,j}} $ and ${P_{teacher,j}}$ are values of learner $ i $ and teacher in dimension $ j $, $P_{_{i,j}}^{(1)}$ is the value updated by $ {P_{i,j}} $, ${M_j}$ is the mean value of all learners in dimension $ j $, ${r_i}$ is a random number among $ [0,1] $, and $ {T_F} = {\rm round}[1 + {\rm rand}(0,1)\{ 2 - 1\} ] $ is a teaching factor. If $P_{_{i,j}}^{(1)}$ is better than ${P_{i,j}}$, it is then accepted by the learner population.

\subsection{Learner phase}
In this phase, the learner ${P_m}$ ($ m=1,2,…,Np $) learns knowledge from another learner ${P_l}$ ($ l=1,2,…,Np $) selected randomly. If ${P_l}$ performs better, ${P_m}$ will move toward it, otherwise ${P_m}$ moves away from it. The learning process can be described as follows
\begin{equation}
P_{_{m,j}}^{(2)} = \left\{ {\begin{array}{*{20}{l}}
	{{P_{m,j}} + {r_m}({P_{m,j}} - {P_{l,j}})\;\;\;if\;f({P_m}) < f({P_l})}\\
	{{P_{m,j}} + {r_m}({P_{l,j}} - {P_{m,j}})\;\;\;otherwise}
	\end{array}} \right.
\label{eq2}
\end{equation}
where ${r_m}$ is a random number among $ [0, 1] $, and $P_{_{m,j}}^{(2)}$ is the updated learner of ${P_{m,j}}$. If the fitness value of $P_{_{m,j}}^{(2)}$ is better, ${P_{m,j}}$ will be replaced by $P_{_{m,j}}^{(2)}$.

\section{Achievable performance and PID tuning}
The CPA problem of PID aims to find the achievable performance measured by minimal output variance (MOV). This section first describes the calculation method of the achievable performance proposed in the literature \cite{ko_pid_2004}, and the non-convex optimization of CPA for the single-loop case is represented. Then this method is extended to the PI/P cascade control, and CPA for the PI/P cascade control is formulated. Finally, a multi-objective function taking into account IAE and MOV simultaneously is proposed to tune the PID control for better disturbance rejection.

\subsection{Achievable performance of single-loop case}
\begin{figure}[htbp]
	\centering
	\includegraphics[width=10cm, height=3.4cm]{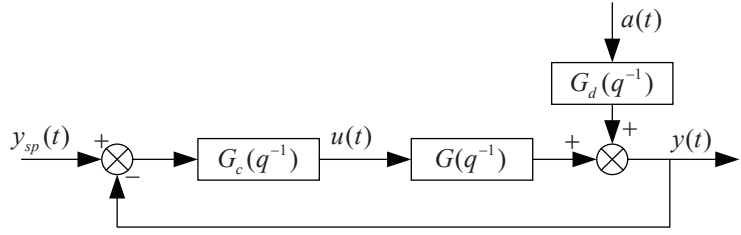}
	\caption{A single closed-loop control system.}
	\label{fig:fig1}
\end{figure}

A typical single-input-single-output (SISO) control system is shown in Figure \ref{fig:fig1}, where $ t $, $ a(t) $, $ u(t) $ and $ y(t) $ represent sampling interval, zero mean white noise, manipulated variable and controlled output, respectively. The output of this system can be written as
\begin{equation}
y(t) = G({q^{ - 1}})u(t) + {G_d}({q^{ - 1}})a(t)
\label{eq3}
\end{equation}
where $G({q^{ - 1}})$ and ${G_d}({q^{ - 1}})$ denote the process and disturbance transfer function, and ${q^{ - 1}}$ is the backward shift operator. It is assumed that $G({q^{ - 1}})$ and ${G_d}({q^{ - 1}})$ are stable, minimum-phase and causal. 

It is further assumed that there is no setpoint change, i.e., ${y_{sp}}(t) = 0$. The output of the system can be described as follows
\begin{equation}
y(t) =  - y(t){G_c}({q^{ - 1}})G({q^{ - 1}}) + a(t){G_d}({q^{ - 1}})
\label{eq4}
\end{equation}
When the structure of the controller is restricted to PID described as follows
\begin{equation}
{G_c}({q^{ - 1}}) = \frac{{{k_1} + {k_2}{q^{ - 1}} + {k_3}{q^{ - 2}}}}{{1 - {q^{ - 1}}}}
\label{eq5}
\end{equation}
where ${k_1} = {k_P} + {k_I} + {k_D}$, ${k_2} =  - ({k_P} + 2{k_D})$ and ${k_3} = {k_D}$. ${k_P}$, ${k_I}$ and ${k_D}$ represent proportional, integral and derivative gain, respectively. And if only a single shock $a(0)$ is introduced to the system, according to the convolution theorem, the calculation of output sequence $\bar y = {[y(0),y(1), \ldots ,y(n)]^T}$ is expressed as follows
\begin{equation}
\bar y =  - {k_1}{I_m}\bar y - {k_2}F{I_m}\bar y - {k_3}{F^2}{I_m}\bar y + \bar na(0)
\label{eq6}
\end{equation}
where $\bar n = {[{g_d}(0),{g_d}(1), \ldots ,{g_d}(n)]^T}$ is the impulse response of the disturbance model, $ F $ is the forward shift matrix and ${I_m}$ is the matrix consist of the impulse response $\bar g = {[g(1),g(2), \ldots ,g(n)]^T}$ of the process model, i.e.,
\begin{center}
$ F = {\left[ {\begin{array}{*{20}{c}}
		0&{}&{}&0\\
		1& \ddots &{}&{}\\
		{}& \ddots & \ddots &{}\\
		0&{}&1&0
		\end{array}} \right]_{(n + 1) \times (n + 1)}} $,
$ {I_m} = \left[ {\begin{array}{*{20}{c}}
		0&0&0& \cdots &0\\
		{g(1)}&0&0& \cdots &0\\
		{g(2)}&{g(1)}&0& \cdots &0\\
		\vdots & \vdots & \vdots & \ddots & \vdots \\
		{g(n)}&{g(n - 1)}&{g(n - 2)}& \cdots &0
		\end{array}} \right] $
\end{center}
The output sequence can be set as
\begin{equation}
\bar y = {(I + {k_1}{I_m} + {k_2}F{I_m} + {k_3}{F^2}{I_m})^{ - 1}}\bar na(0) = \varphi a(0)
\label{eq7}
\end{equation}
where $\varphi  = {[\varphi (1),\varphi (2), \ldots ,\varphi (n)]^T}$ is the impulse response of the closed-loop model, and the output variance can be calculated as \cite{ko_pid_2004}
\begin{equation}
\sigma _y^2 = {\varphi ^T}\varphi \sigma _a^2
\label{eq8}
\end{equation}
where $\sigma _a^2$ is the variance of disturbance. 
Therefore, the CPA problem of PID control is described as follows
\begin{equation}
{J_1} = \min \;{f_1}({k_1},{k_2},{k_3}) = \mathop {\min }\limits_{{k_1},{k_2},{k_3}} \;\sigma _y^2 = \mathop {\min }\limits_{{k_1},{k_2},{k_3}} \;{\varphi ^T}\varphi \sigma _a^2
\label{eq9}
\end{equation}
The impulse response with finite length p (i.e., ${\varphi _p} = {[\varphi (1),\varphi (2),...,\varphi (p)]^T}$) is utilized to approximate the output variance, and the non-convex problem can be redescribed as follows
\begin{equation}
{J_1} = {\min}\;f_1({k_1},{k_2},{k_3}) \approx \mathop {\min }\limits_{{k_1},{k_2},{k_3}} \varphi _p^T{\varphi _p}\sigma _a^2
\label{eq10}
\end{equation}

\subsection{Achievable performance of PI/P cascade control}
\begin{figure}[htbp]
	\centering
	\includegraphics[width=15.6cm, height=3.4cm]{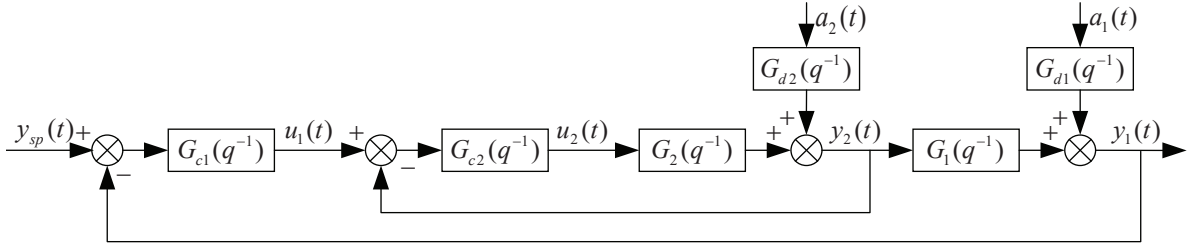}
	\caption{A cascade control system.}
	\label{fig:fig2}
\end{figure}

Figure \ref{fig:fig2} shows a cascade control system with the outer loop model ${G_1}({q^{ - 1}})$ and inner loop model ${G_2}({q^{ - 1}})$. ${a_1}(t)$ and ${a_2}(t)$ are disturbances in the outer and in the inner loop, respectively. The disturbance models are ${G_{d1}}({q^{ - 1}})$ and ${G_{d2}}({q^{ - 1}})$. When the setpoint ${y_{sp}}(t) = 0$, and the structures of the primary controller ${G_{c1}}$ and the secondary controller ${G_{c2}}$ are restricted to PI and P, respectively, i.e.,
\begin{equation}
{G_{c1}} = \frac{{{k_4} + {k_5}{q^{ - 1}}}}{{1 - {q^{ - 1}}}},{G_{c2}} = {k_6}
\label{eq11}
\end{equation}
The outputs of the outer loop ${y_1}(t)$ and the inner loop ${y_2}(t)$ are
\begin{equation}
\begin{array}{l}
{y_1}(t) = {y_2}(t){G_1}({q^{ - 1}}) + {a_1}(t){G_{d1}}({q^{ - 1}})\\
{y_2}(t) = {k_6}[ - {y_1}(t)\frac{{{k_4} + {k_5}{q^{ - 1}}}}{{1 - {q^{ - 1}}}} - {y_2}(t)]{G_2}({q^{ - 1}}) + {a_2}(t){G_{d2}}({q^{ - 1}})
\end{array}
\label{eq12}
\end{equation}

When the system is only influenced by the initial shocks of the disturbances ${a_1}(0)$ and ${a_2}(0)$, the output vectors are formulated as follows
\begin{equation}
\begin{array}{l}
{{\bar y}_1} = {I_{m1}}{{\bar y}_2} + {{\bar n}_1}{a_1}(0)\\
{{\bar y}_2} =  - {k_4}{k_6}{S_2}{{\bar y}_1} - {k_5}{k_6}F{S_2}{{\bar y}_1} - {k_6}{I_{m2}}{{\bar y}_2} + {{\bar n}_2}{a_2}(0)
\end{array}
\label{eq13}
\end{equation}
where $ {\bar y_i} = {[{y_i}(0),{y_i}(1), \ldots ,{y_i}(n)]^T}(i = 1,2) $, $ {\bar n_i} = {[{g_{di}}(0),{g_{di}}(1), \ldots ,{g_{di}}(n)]^T}(i = 1,2) $ is the impulse response of the disturbances, ${I_{mi}}(i = 1,2)$ are the matrices consist of the impulse response ${\bar g_i} = {[{g_i}(1),{g_i}(2), \ldots ,{g_i}(n)]^T}(i = 1,2)$ of the process models, and $ {S_2} $ is the matrix consist of the step response ${\bar s_2} = {[{s_2}(1),{s_2}(2), \ldots ,{s_2}(n)]^T}$ of the process model of inner loop.

The output vector can be expressed in the following form 
\begin{equation}
{\bar y_1} = {\varphi _1}{a_1}(0) + {\varphi _2}{a_2}(0)
\label{eq14}
\end{equation}
where
\begin{equation}
\begin{array}{l}
{\varphi _1} = {(I + W)^{ - 1}}{{\bar n}_1}\\
{\varphi _2} = {(I + W)^{ - 1}}[{I_{m1}}{(I + {k_6}{I_{m2}})^{ - 1}}]{{\bar n}_2}\\
W = {k_4}{k_6}{I_{m1}}{(I + {k_6}{I_{m2}})^{ - 1}}{S_2} + {k_5}{k_6}{I_{m1}}{(I + {k_6}{I_{m2}})^{ - 1}}F{S_2}
\end{array}
\label{eq15}
\end{equation}
The variance of the output is 
\begin{equation}
\sigma _{{y_1}}^2 = \varphi _1^T{\varphi _1}\sigma _{{a_1}}^2 + \varphi _2^T{\varphi _2}\sigma _{{a_2}}^2 + 2\varphi _1^T{\varphi _2}{\sigma _{{a_1}}}{\sigma _{{a_2}}}
\label{eq16}
\end{equation}
The CPA problem of the PI/P cascade control can be described as follows
\begin{equation}
{J_2} = \min \;{f_2}({k_4},{k_5},{k_6}) = \mathop {\min }\limits_{{k_4},{k_5},{k_6}} \;\sigma _{y1}^2 = \mathop {\min }\limits_{{k_4},{k_5},{k_6}} \;\varphi _1^T{\varphi _1}\sigma _{{a_1}}^2 + \varphi _2^T{\varphi _2}\sigma _{{a_2}}^2 + 2\varphi _1^T{\varphi _2}{\sigma _{{a_1}}}{\sigma _{{a_2}}}
\label{eq17}
\end{equation}

\subsection{PID tuning based on a new multi-objective function}
A single objective function can’t optimize all performance criteria of a control system at the same time. For example, the most commonly used objective function IAE can’t optimize the overshoot and the settling time at the same time because they conflict with each other. Multi-objective optimization is a technique to solve optimization problems that involve two or more conflicting object functions \cite{fonseca_multiobjective_1998}. Unlike single objective optimization with only one “best solution”, it always has a set of alternative optima. These solutions are called Pareto optimal set, and a decision-making process is needed to select an appropriate compromise solution from the set.

The performance of the stochastic disturbance rejection is one of the most important criteria for a control system. However, few studies have designed objective functions taking into account it. Moreover, this criterion is always in conflict with other performance criteria such as overshoot and settling time. Therefore, most of the existing tuning methods can’t find the best solution for better stochastic disturbance rejection. To solve this problem, a new multi-objective function considering both MOV and IAE is designed, which is described as follows
\begin{equation}
{J_3} = \min {f_3}({k_{PID}}) = \mathop {\min }\limits_{{k_{PID}}} (\int_0^\infty  {\left| {e(t)} \right|dt}  + \rho \sigma _y^2)
\label{eq18}
\end{equation}

where ${k_{PID}}$ is the PID parameter, $e(t) = {y_{sp}}(t) - y(t)$ is the error between the setpoint and the output, $\sigma _y^2$ is the output variance, and $\rho $ is a weight. It should be noted that the calculation of $ IAE=\int_0^\infty  {\left| {e(t)} \right|dt} $ is in the case that the system is influenced by the setpoint but not the disturbance, and the calculation of output variance is just the opposite case.

By adjusting the weight $\rho $ in a proper range, this function can compromise other performance criteria related to IAE to improve the stochastic disturbance rejection. Owing to the fact that IAE is always much larger than the output variance, a relatively large weight is necessary. Otherwise, too small a weight can’t attain a better performance of disturbance rejection. The performance criteria such as overshoot and settling time mainly concern the initial stage of step response, but disturbance rejection concerns the steady state stage. Therefore, combining this tuning method with the multi-stage PID tuning strategy can resolve the contradiction between disturbance rejection and other performance criteria, i.e., the weight is set to 0 or a small value in the initial stage and set to a relatively large value in the steady state stage.

\subsection{The steps of algorithm}
The CPA problem described by (\ref{eq10}) and the tuning problem described by (\ref{eq18}) are solved by the TLBO algorithm, the steps of which are present in Algorithm \ref{tab:algorithm1}.
\begin{algorithm}[!h]
	\caption{Steps of the TLBO algorithm.}
	\label{tab:algorithm1}
	Ensure a proper range for controller parameters\;
	Randomly generate an initial learner population in the search space\;
	Set $G = 0$\;
	\While{(the termination criterion is not met)}
	{
		Select the teacher $ P_{_{teacher,j}}^G $\;
		\For {$ i = 1;{\rm{ }}i \le Np;{\rm{ }}i++ $}
		{
			Update the position of the learner ${P_i}$ according to (\ref{eq1})\;
			if $ f_j(P_i^{(1)}) < f({P_i})(j=1,3) $ then\\
			Replace ${P_i}$ with $P_i^{(1)}$\;
		}
		G=G+1\;
		\For {$ i = 1;{\rm{ }}i \le Np;{\rm{ }}i +  +  $}
		{
			Update the position of the learner according to (\ref{eq2})\;
			if $ f_j(P_i^{(2)}) < f({P_i})(j=1,3) $ then\\
			Replace ${P_i}$ with $P_i^{(2)}$\;
		}
		G=G+1\;
	}	
\end{algorithm}

\section{Simulation examples}
\subsection{CPA of single-loop case}
In this section, ten benchmark problems (as shown in Table \ref{tab:PID benchmark}) adopted from the literature \cite{shahni_assessment_2011} are used to verify the excellent performance of the algorithm in solving the non-convex problem. To reduce the error of approximation to obtain an accurate MOV, the length of the impulse response is selected as $ p=8d $ \cite{shahni_assessment_2011}, where $ d $ is the time delay of the process model. After some experiments, the parameters chosen for the algorithm are as follows: the number of learners is $ Np=20 $, the search space of PID parameters is set as $ [-50, 50] $, and the termination criterion is designed as $f(P_{_{teacher,j}}^G) - f(P_{_{teacher,j}}^{G - 20}) < {10^{ - 7}}$, where $P_{_{teacher,j}}^G$ is the teacher at iteration $ G $. All experiments were demonstrated 30 times independently to test the stability of the algorithm, and were run on Matlab R2017a on Intel(R) Core(TM) i5-4460 CPU @ 3.20GHz with 12GB RAM".

\renewcommand\arraystretch{1}
\begin{table}[!h]
	\centering\small
	\begin{threeparttable}
		\caption{Benchmark problems of PID performance assessment \cite{shahni_assessment_2011}.}\label{tab:PID benchmark}
		\begin{tabular}{ccc}
			\hline	
			Example &$G$ &$G_d$
			\\
			\hline
			$1$ &$\frac{{0.2{q^{ - 5}}}}{{1 - 0.8{q^{ - 1}}}}$ &$\frac{1}{{(1 - {q^{ - 1}})(1 + 0.4{q^{ - 1}})}}$ 
			\\
			$2$ &$\frac{{0.08919{q^{ - 12}}}}{{1 - 0.8669{q^{ - 1}}}}$ &$\frac{{0.08919}}{{1 - 0.8669{q^{ - 1}}}}$ 
			\\
			$3$ &$\frac{{0.5108{q^{ - 28}}}}{{1 - 0.9604{q^{ - 1}}}}$ &$\frac{{0.5108}}{{1 - 0.9604{q^{ - 1}}}}$ 
			\\
			$4$ &$\frac{{{q^{ - 6}}}}{{1 - 0.8{q^{ - 1}}}}$ &$\frac{{1 + 0.6{q^{ - 1}}}}{{(1 - 0.5{q^{ - 1}})(1 - 0.6{q^{ - 1}})(1 + 0.7{q^{ - 1}})}}$ 
			\\
			$5$ &$\frac{{{q^{ - 6}}}}{{1 - 0.8{q^{ - 1}}}}$ &$\frac{{1 - 0.2{q^{ - 1}}}}{{(1 - {q^{ - 1}})(1 - 0.3{q^{ - 1}})(1 + 0.4{q^{ - 1}})(1 - 0.5{q^{ - 1}})}}$ 
			\\
			$6$ &$\frac{{{q^{ - 6}}}}{{1 - 0.8{q^{ - 1}}}}$ &$\frac{{1 + 0.6{q^{ - 1}}}}{{(1 - {q^{ - 1}})(1 - 0.5{q^{ - 1}})(1 - 0.6{q^{ - 1}})(1 + 0.7{q^{ - 1}})}}$ 
			\\
			$7$ &$\frac{{0.1{q^{ - 5}}}}{{1 - 0.8{q^{ - 1}}}}$ &$\frac{{0.1}}{{(1 - {q^{ - 1}})(1 - 0.3{q^{ - 1}})(1 - 0.6{q^{ - 1}})}}$ 
			\\
			$8$ &$\frac{{0.1{q^{ - 3}}}}{{1 - 0.8{q^{ - 1}}}}$ &$\frac{1}{{1 - {q^{ - 1}}}}$ 
			\\
			$9$ &$\frac{{0.1{q^{ - 6}}}}{{1 - 0.8{q^{ - 1}}}}$ &$\frac{{0.1}}{{(1 - {q^{ - 1}})(1 - 0.7{q^{ - 1}})}}$ 
			\\
			$10$ &$\frac{{0.1{q^{ - 3}}}}{{1 - 0.8{q^{ - 1}}}}$ &$\frac{{\sqrt {0.001} }}{{(1 - {q^{ - 1}})(1 + 0.2{q^{ - 1}})}}$ 
			\\
			\hline
		\end{tabular}
	\end{threeparttable}
\end{table}

The results of TLBO and the best known results of the reference \cite{shahni_assessment_2011,fu_pid_2012,shahni_rapid_2019} are shown in Table \ref{tab:benchmark_result}, where “MV” is the minimum variance benchmark, “BKMOV” is the best known results, and “Mean”, “Std”, “Worst” and “Time” are the mean, the standard derivation, the worst and the mean calculation time of 30 runs, respectively. It shows that TLBO has better MOV on problems 3, 4, 5, 6 and 9, and has the same MOV on other problems. Particularly, the calculation time of TLBO is less than one second on most problems. The mean and standard derivation of 30 runs of the MOV-related PID parameters are shown in Table \ref{tab:PID_parameters}. It reveals that the TLBO algorithm can solve the non-convex problem with accurate estimation, high efficiency and good stability.

\renewcommand\arraystretch{0.75}
\begin{table}[H]
	\centering
	\begin{threeparttable}
		\caption{Best known MOV (BKMOV) and results of TLBO.}
		\label{tab:benchmark_result}
			\begin{tabular}{lllllll}
			\hline
			\small Example&\small MV&\small BKMOV&\small Mean&\small Std&\small Worst&\small Time(s)\\
			\hline
			\small 1     & \footnotesize 2.9427  & \footnotesize 3.0728  & \footnotesize 3.0728  & \footnotesize 3.36E-10 & \footnotesize 3.0728  & \footnotesize 0.3106  \\
			\small 2     & \footnotesize 0.0306  & \footnotesize 0.0310  & \footnotesize 0.0310  & \footnotesize 2.15E-11 & \footnotesize 0.0310  & \footnotesize 0.7524  \\
			\small 3     & \footnotesize 3.0112  & \footnotesize 3.0238  & \footnotesize \textBF{3.0232 } & \footnotesize 5.16E-10 & \footnotesize 3.0232  & \footnotesize 3.6852  \\
			\small 4     & \footnotesize 3.4004  & \footnotesize 3.4065  & \footnotesize \textBF{3.4064 } & \footnotesize 4.94E-09 & \footnotesize 3.4064  & \footnotesize 0.3624  \\
			\small 5     & \footnotesize 11.9528  & \footnotesize 13.8076  & \footnotesize \textBF{13.8068 } & \footnotesize 5.18E-07 & \footnotesize 13.8068  & \footnotesize 0.3800  \\
			\small 6     & \footnotesize 58.3406  & \footnotesize 87.7377  & \footnotesize \textBF{87.7069 } & \footnotesize 7.88E-10 & \footnotesize 87.7069  & \footnotesize 0.4128  \\
			\small 7     & \footnotesize 0.2978  & \footnotesize 0.4246  & \footnotesize 0.4246  & \footnotesize 5.36E-08 & \footnotesize 0.4246  & \footnotesize 0.2691  \\
			\small 8     & \footnotesize 3.0000  & \footnotesize 3.2032  & \footnotesize 3.2032  & \footnotesize 3.40E-08 & \footnotesize 3.2032  & \footnotesize 0.1900  \\
			\small 9     & \footnotesize 0.3144  & \footnotesize 0.4268  & \footnotesize \textBF{0.4267 } & \footnotesize 2.50E-09 & \footnotesize 0.4267  & \footnotesize 0.3395  \\
			\small 10    & \footnotesize 0.0023  & \footnotesize 0.0024  & \footnotesize 0.0024  & \footnotesize 2.41E-10 & \footnotesize 0.0024  & \footnotesize 0.1436  \\
			\hline	
		\end{tabular}
		\footnotesize {$^*$The bolder ones mean the best results.}\\				
	\end{threeparttable}
\end{table}

\renewcommand\arraystretch{0.65}
\begin{table}[H]
	\centering
	\begin{threeparttable}
		\caption{The Mean and Std of PID parameters$([k_1,k_2,k_3])$.}
		\label{tab:PID_parameters}
		\begin{tabular}{lll}
			\hline
			\small Example & \small Mean  & \small Std \\
			\hline
			\small 1     & \footnotesize [2.8408, -4.4059, 1.7486] & \footnotesize [1.51E-05, 9.22E-05, 4.53E-05] \\
			\small 2     & \footnotesize [1.8236, -3.3531, 1.5299] & \footnotesize [1.31E-04, 6.84E-04, 3.12E-04] \\
			\small 3     & \footnotesize [0.4989, -0.9663, 0.4674] & \footnotesize [1.17E-05, 3.71E-05, 2.03E-05] \\
			\small 4     & \footnotesize [0.1354, -0.2523, 0.1170] & \footnotesize [8.00E-06, 1.47E-05, 7.19E-06] \\
			\small 5     & \footnotesize [0.7241, -1.2058, 0.5178] & \footnotesize [1.25E-05, 3.34E-06, 1.82E-06] \\
			\small 6     & \footnotesize [0.8327, -1.4003, 0.6094] & \footnotesize [5.00E-07, 7.67E-06, 4.33E-06] \\
			\small 7     & \footnotesize [8.0941, -13.1891, 5.5927] & \footnotesize [7.27E-04, 4.69E-04, 2.55E-04] \\
			\small 8     & \footnotesize [6.5338, -9.2379, 3.3583] & \footnotesize [3.74E-05, 1.79E-04, 1.16E-04] \\
			\small 9     & \footnotesize [8.2318, -13.7793, 5.9701] & \footnotesize [1.00E-04, 2.51E-04, 1.45E-04] \\
			\small 10    & \footnotesize [6.1676, -8.5741, 3.0332] & \footnotesize [5.73E-04, 1.35E-03, 7.63E-04] \\
			\hline	
		\end{tabular}
	\end{threeparttable}
\end{table}

\subsection{Tuning of single-loop case}
\begin{figure}[H]
	\centering
	\includegraphics[width=12cm, height=6.6cm]{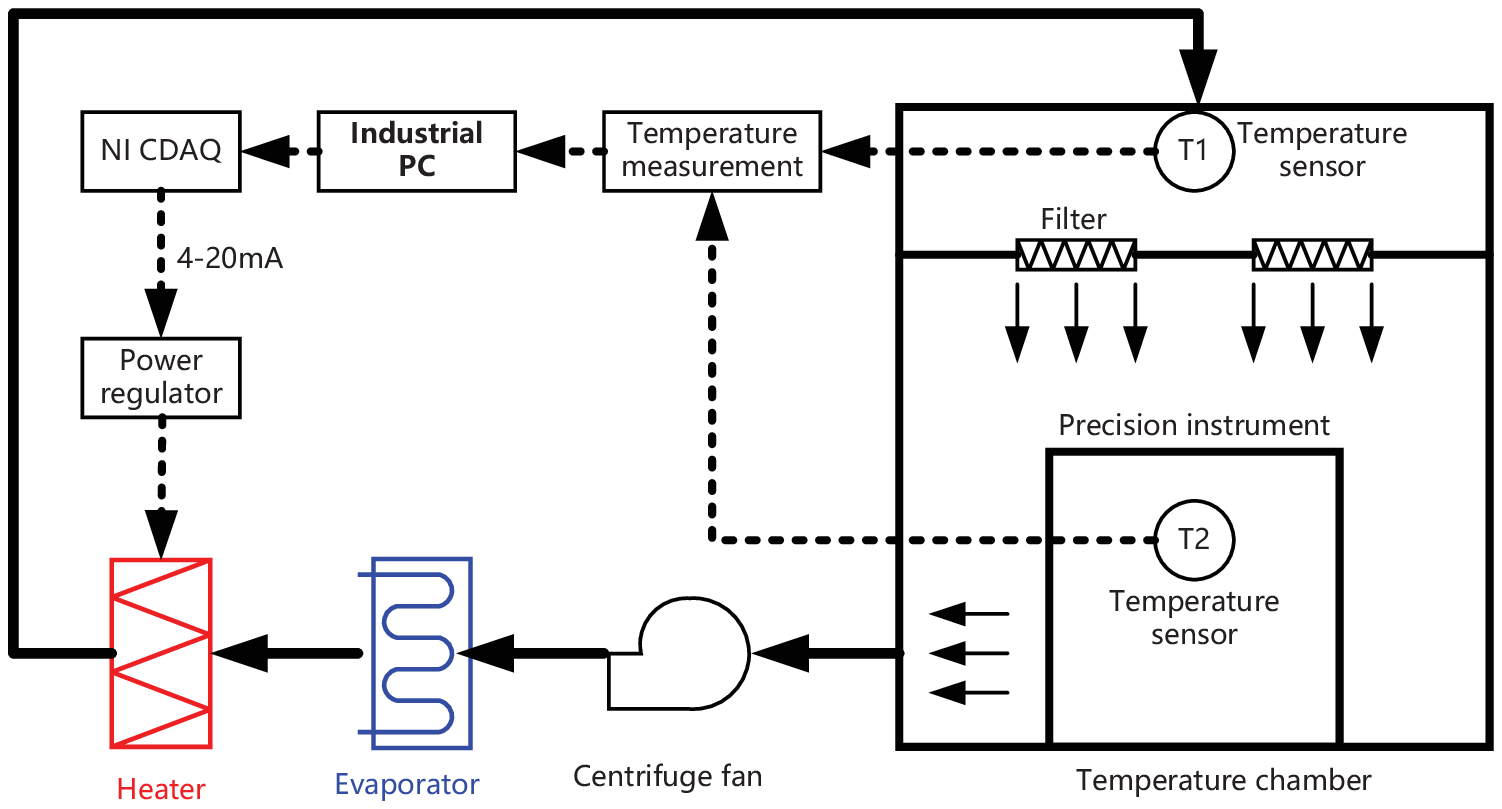}
	\caption{Schematic of the air temperature control system.}
	\label{fig:tempctrlsys}
\end{figure}

The tuning method based on the multi-objective optimization for single-loop PID is applied to a high-precision air temperature control system, which provides an environment with high temperature stability for precision instruments such as laser interferometers and lithography tools \cite{zhao_optimization_2010}. As shown in Figure \ref{fig:tempctrlsys}, the temperature control system aims to supply air with high temperature stability to the temperature chamber, that is to maintain the temperature of the point “T1” measured by a thermistor. The temperature is controlled by a pipe heater, the power of which is adjusted by a power regulator receiving 4-20mA current signal. 

The input of the single-loop temperature control system is the current (mA), and the output is the temperature ($^ \circ C$) of the point “T1”. A step test is implemented to identify a first-order plus dead time (FOPDT) model as follows
\begin{equation}
G(s) = \frac{{y(s)}}{{u(s)}} = \frac{{0.{\rm{39}}}}{{{\rm{90}}{\rm{.28}}s + 1}}{e^{ - {\rm{31}}{\rm{.98}}s}}
\label{eq19}
\end{equation}
This model is discretized with 10 seconds as the sampling time, and the discrete form is:
\begin{equation}
G(q^{-1}) = \frac{{0.04{\rm{13}}{q^{ - 4}}}}{{1 - 0.895{\rm{2}}{q^{ - 1}}}}
\label{eq20}
\end{equation}
The disturbance model for the process is simulated as \cite{fang_g113-lqg_2017}:
\begin{equation}
h({q^{ - 1}}) = \frac{{0.2}}{{1 - 0.8951{q^{ - 1}}}}
\label{eq21}
\end{equation}
with the variance of the noise is $\sigma _a^2 = {10^{ - 5}}$. Therefore, the model of the process is
\begin{equation}
y = \frac{1}{1-q^{-1}}\left[ {\frac{{0.04{\rm{13}}{q^{ - 4}}}}{{1 - 0.895{\rm{2}}{q^{ - 1}}}}(1-q^{-1})u + \frac{{0.2}}{{1 - 0.895{\rm{2}}{q^{ - 1}}}}a} \right]
\label{eq22}
\end{equation}

\begin{figure}[H]
	\centering
	\includegraphics[width=8cm, height=6cm]{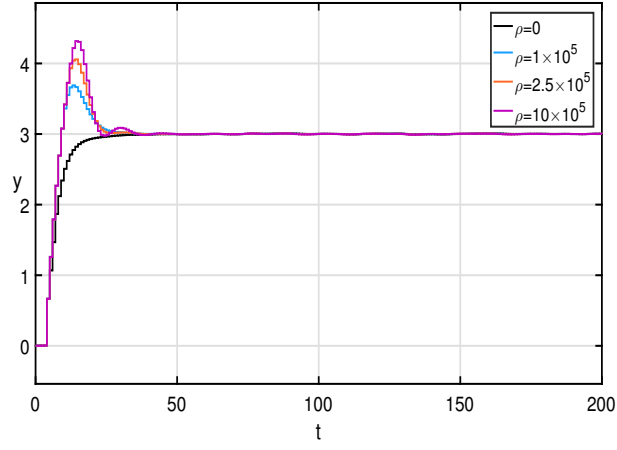}
	\caption{Step response of single-loop case.}
	\label{fig:step_single}
\end{figure}

The effectiveness of the tuning method is verified by comparing the results of four weights. The output variance is presented in Table \ref{tab:single}, and the response curves under the step change of the setpoint are shown in Figure \ref{fig:step_single}. It reveals that adjusting the weight can improve the stability of temperature control but will lead to a large overshoot in the initial stage. To solve this problem, a relatively small weight can be used in the initial stage.

\renewcommand\arraystretch{0.75}
\begin{table}[!htbp]
	\centering
	\begin{threeparttable}
		\caption{Output variance of the single-loop case.}
		\label{tab:single}
		\begin{tabular}{ccc}
			\hline
			\small $\rho$ ($\times10^5$) & \small $[k_1,k_2,k_3]$ & \small $\sigma _y^2$ ($\times10^{-5}$) \\
			\hline
			\small 0     & \footnotesize [5.3333, -6.8756, 1.8693] & \footnotesize 7.7624 \\
			\small 1     & \footnotesize [7.9520, -10.2099, 2.8804] & \footnotesize 4.0747 \\
			\small 2.5     & \footnotesize [9.5647, -12.4166, 3.6362] & \footnotesize 3.2726 \\
			\small 10     & \footnotesize [23.1165, -35.5929, 14.4531] & \footnotesize 2.6432 \\
			\hline	
		\end{tabular}
	\end{threeparttable}
\end{table}

\subsection{Tuning of PI/P cascade control}
\begin{figure}[H]
	\centering
	\includegraphics[width=9cm, height=5.5cm]{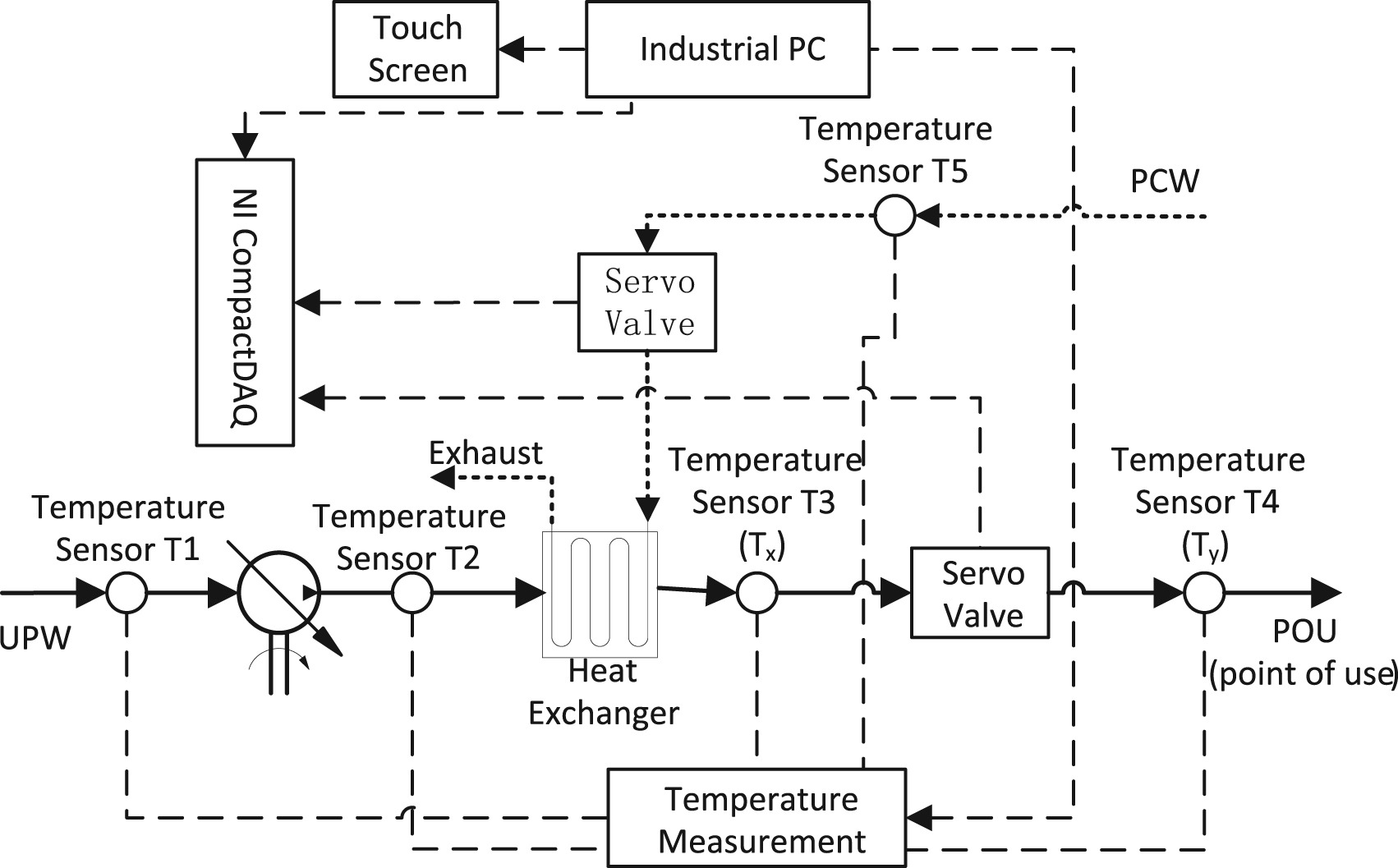}
	\caption{Schematic of the immersion liquid temperature control system \cite{li_high_2017}.}
	\label{fig:immersion}
\end{figure}

The temperature control system of immersion liquid in immersion lithography (as shown in Figure \ref{fig:immersion}) adopted from the literature \cite{li_high_2017} is tested to investigate the tuning of the PI/P cascade control based on the multi-objective function. The controlled variable is the temperature of immersion liquid ``$ \rm T_y $'', and the manipulated variable is the flow rate of cooling water (PCW) controlled by a valve. Since the pipe between ``T3'' and ``T4'' is long, a cascade control is used to improve the disturbance rejection, and the sensor of the inner loop is ``T3''. The models of the outer loop and inner loop of this system are described as follows
\begin{equation}
{G_1}(s) = \frac{{1.0092}}{{1 + 138.06s}}{e^{ - 35.75s}},{G_2}(s) = \frac{{ - 1.3361}}{{1 + 11.834s}}{e^{ - 11.63s}}
\label{eq23}
\end{equation}
The discrete models with the sampling time of 6s are
\begin{equation}
{G_1}({q^{ - 1}}) = \frac{{0.04292}}{{1 - 0.9575{q^{ - 1}}}}{q^{ - 7}},{G_2}({q^{ - 1}}) = \frac{{ - 0.5314}}{{1 - 0.6023{q^{ - 1}}}}{q^{ - 3}}
\label{eq24}
\end{equation}
The disturbance models are simulated as
\begin{equation}
{G_{d1}}({q^{ - 1}}) = \frac{1}{{1 - 0.9575{q^{ - 1}}}},{G_{d2}}({q^{ - 1}}) = \frac{1}{{1 - 0.6023{q^{ - 1}}}}
\label{eq25}
\end{equation}
and the variances of the disturbances are set as $\sigma _{{a_1}}^2 = 0.0005,\sigma _{{a_2}}^2 = 0.005$.

\begin{figure}[H]
	\centering
	\includegraphics[width=8cm, height=6cm]{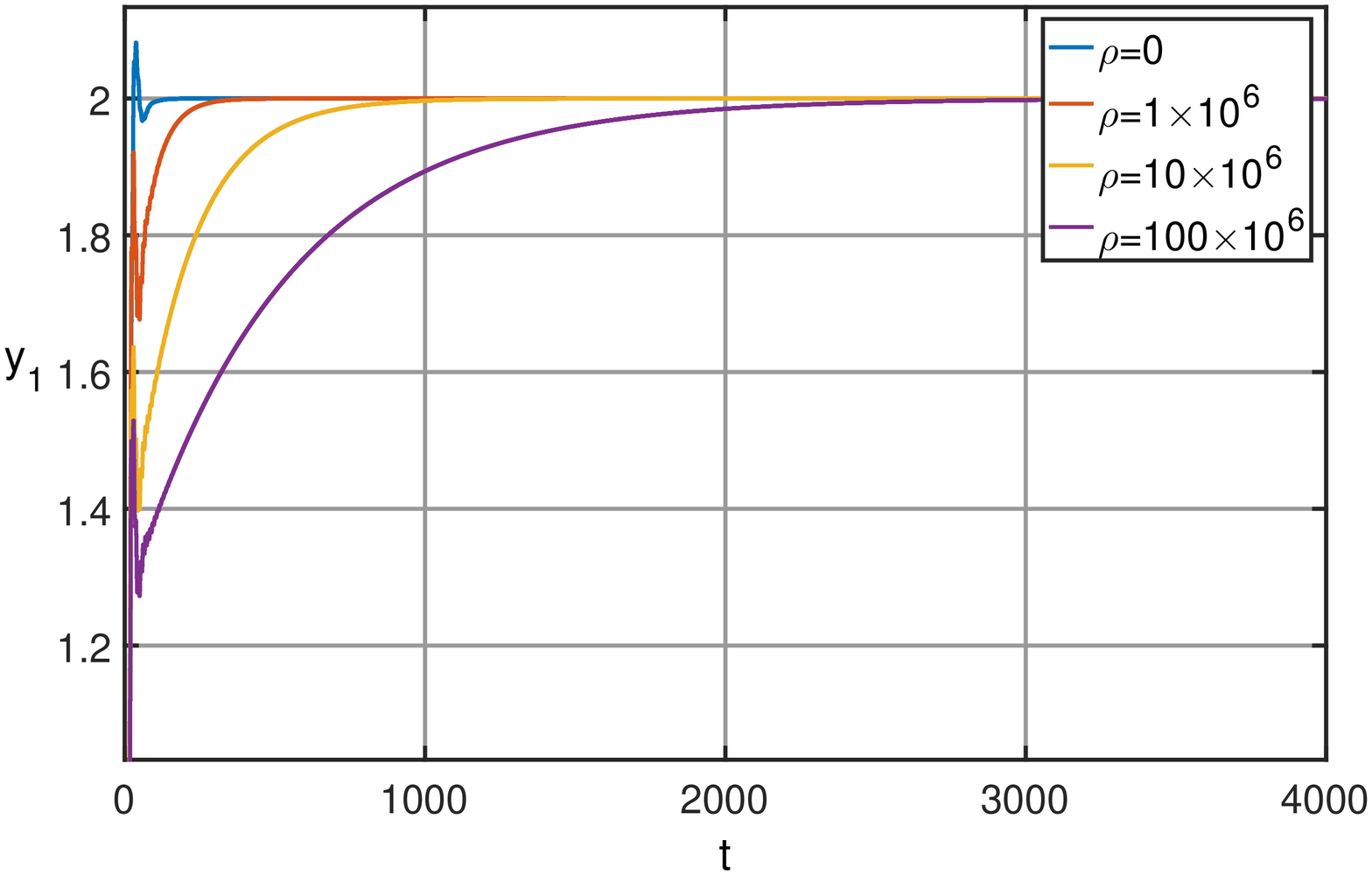}
	\caption{Step response of PI/P cascade control.}
	\label{fig:step_cascade}
\end{figure}

The test results of four weights are shown in Table \ref{tab:cascade} and Figure \ref{fig:step_cascade}. It indicates that a larger weight relates to a smaller output variance, but the settling time is longer. To solve this conflict, a relatively smaller weight can be used to stabilize the system quickly, and then a larger weight is utilized to improve the disturbance rejection to attain a better performance of temperature control. 

\renewcommand\arraystretch{0.75}
\begin{table}[!htbp]
	\centering
	\begin{threeparttable}
		\caption{Output variance of PI/P cascade control.}
		\label{tab:cascade}
		\begin{tabular}{ccc}
			\hline
			\small $\rho$ ($\times10^6$) & \small $[k_4,k_5,k_6]$ & \small $\sigma_{y_1}^2$ ($\times10^{-4}$) \\
			\hline
			\small 0     & \footnotesize [2.7638, -2.6554, -0.8436] & \footnotesize 6.0551 \\
			\small 1     & \footnotesize [3.0563, -2.9922, -0.9631] & \footnotesize 5.3566 \\
			\small 10     & \footnotesize [2.8715, -2.8482, -1.0054] & \footnotesize 4.9421 \\
			\small 100     & \footnotesize [2.9088, -2.8420, -0.9538] & \footnotesize 4.8117 \\
			\hline		
		\end{tabular}
	\end{threeparttable}
\end{table}

\section{Conclusion}
This paper proposes a multi-objective function considering both IAE and MOV for PID tuning to improve the stochastic disturbance rejection. The TLBO algorithm is employed to solve the multi-objective optimization problem and the CPA related non-convex problem. Furthermore, the tuning method and CPA are extended to the PI/P cascade control. The TLBO algorithm was tested on ten numerical CPA examples adopted from the literature. The results show that in most examples, TLBO obtains better MOV than the existing methods, and the calculation time is less than one second. The tuning method is applied to a single-loop air temperature control system and a cascade immersion liquid temperature control. The results verify that this method has the ability to improve the disturbance rejection for better performance of temperature control. Combined with the multi-stage PID tuning strategy, this method can resolve the contradiction between the stochastic disturbance rejection and other performance criteria such as the overshoot and settling time.

\section{Acknowledgements}

\bibliography{mybibfile}

\end{document}